\documentclass[onecolumn,final]{IEEEtran}
\usepackage{amsmath,amssymb}
\usepackage{graphicx}
\newcommand{\expectl}[2]{{\mathop{{\mathbb E}}}_{#1}\left[#2\right]}
\newcommand{\expect}[1]{{\mathop{{\mathbb E}}}\left[#1\right]}
\newcommand{\LB}{\left(}
\newcommand{\RB}{\right)}
\newcommand{\LSB}{\left[}
\newcommand{\RSB}{\right]}
\newcommand{\htp}{^{\sf H}}
\newcommand{\tp}{^{\sf T}}

\newfont{\bbb}{msbm10 scaled 500}

\newfont{\bb}{msbm10 scaled 1100}
\newcommand{\CC}{\mbox{\bb C}}
\newcommand{\RR}{\mbox{\bb R}}

\newcommand{\mv}{{\bf m}}
\newcommand{\nv}{{\bf n}}

\newcommand{\qv}{{\bf q}}

\newcommand{\xv}{{\bf x}}
\newcommand{\yv}{{\bf y}}
\newcommand{\zv}{{\bf z}}
\newcommand{\zerov}{{\bf 0}}
\newcommand{\onev}{{\bf 1}}

\newcommand{\Am}{{\bf A}}

\newcommand{\Dm}{{\bf D}}

\newcommand{\Hm}{{\bf H}}
\newcommand{\Id}{{\bf I}}

\newcommand{\Km}{{\bf K}}

\newcommand{\Rm}{{\bf R}}

\newcommand{\Tm}{{\bf T}}

\newcommand{\Wm}{{\bf W}}
\newcommand{\Vm}{{\bf V}}
\newcommand{\Xm}{{\bf X}}
\newcommand{\Ym}{{\bf Y}}

\newcommand{\Cc}{{\cal C}}

\newcommand{\Kc}{{\cal K}}

\newcommand{\Nc}{{\cal N}}

\newcommand{\Sc}{{\cal S}}

\newcommand{\diag}{{\hbox{diag}}}

\renewcommand{\det}{{\hbox{det}}}
\newcommand{\trace}{{\hbox{tr}}\,}

\renewcommand{\arg}{{\hbox{arg}}}

\renewcommand{\Im}{{\rm Im}}

\newcommand{\defines}{{\,\,\stackrel{\scriptscriptstyle \bigtriangleup}{=}\,\,}}

\newtheorem{theorem}{Theorem}

\newtheorem{remark}{\indent Remark}[section]

\begin{document}
\title{Optimal Channel Training in Uplink Network MIMO Systems}
\author{Jakob~Hoydis, Mari~Kobayashi, and M\'{e}rouane~Debbah
\thanks{This work was supported in part by the French cluster System@tic through the project POSEIDON, by the Agence Nationale de la Recherche through the project ANR-09-VERS0: ECOSCELLS, and by the European Commission through the FP7 project WiMAGIC (www.wimagic.eu). Parts of this work have been presented at the IEEE International Conference on Acoustics, Speech and Signal Processing (ICASSP), 2011.}
\thanks{Copyright (c) 2011 IEEE. Personal use of this material is permitted. However, permission to use this material for any other purposes must be obtained from the IEEE by sending a request to pubs-permissions@ieee.org.}
\thanks{J. Hoydis and  M. Kobayashi are with the Department of Telecommunications, Sup\'{e}lec, 91192 Gif-sur-Yvette, France (e-mail: jakob.hoydis@supelec.fr; mari.kobayashi@supelec.fr).}
\thanks{M. Debbah is with the Alcatel-Lucent Chair on Flexible Radio, Sup\'{e}lec, 91192 Gif-sur-Yvette, France. (e-mail: merouane.debbah@supelec.fr).}
}
\maketitle

\begin{abstract}
We consider a multi-cell frequency-selective fading uplink channel (network MIMO) from $K$ single-antenna user terminals (UTs) to $B$ cooperative base stations (BSs) with $M$ antennas each. The BSs, assumed to be oblivious of the applied codebooks, forward compressed versions of their observations to a central station (CS) via capacity limited backhaul links. The CS jointly decodes the messages from all UTs. Since the BSs and the CS are assumed to have no prior channel state information (CSI), the channel needs to be estimated during its coherence time. Based on a lower bound of the ergodic mutual information, we determine the optimal fraction of the coherence time used for channel training, taking different path losses between the UTs and the BSs into account. We then study how the optimal training length is impacted by the backhaul capacity. Although our analytical results are based on a large system limit, we show by simulations that they provide very accurate approximations for even small system dimensions. 
\end{abstract}
\begin{IEEEkeywords}
 Coordinated Multi-Point (CoMP), network MIMO, multi-cell processing, channel estimation, imperfect channel state information (CSI), random matrix theory. 
\end{IEEEkeywords}

\section{Introduction}
\IEEEPARstart{N}{etwork} MIMO has become the synonym for cooperative communications in the cellular context and is regarded as an important concept to boost the interference limited performance of today's cellular networks. It is often also referred to as multi-cell processing or distributed antenna systems and corresponds to a communication system where multiple base stations (BSs), connected via high speed backhaul links to a central station (CS), \textit{jointly} process data either received over the uplink or transmitted over the downlink. If the BSs could cooperate without any restrictions with regards to the backhaul capacity, processing delay, computing complexity and the availability of channel state information (CSI), the multi-cell interference channel would be transformed into a multiple-access (uplink) or broadcast (downlink) channel without multi-cell interference. This argument motivated the concept of network MIMO and it has been shown in many works, e.g. \cite{Venkatesan2007}, that BS-cooperation has the potential to realize significant gains in throughput and reliability. 

So far, the treatment of multi-cell cooperation in the literature has been either information-theoretic but limited to simple models \cite{Shamai2004,Somekh2007b} or based on simulations to account for more realistic and complex network structures \cite{Marsch2008b,Ramprashad2009,Ramprashad2009a}. The most common and analytically tractable network models are the Wyner model \cite{Hanly1993,wyner94} and the soft hand-off model \cite{Somekh2007a,Simeone2009} which consider cooperation between either two or three adjacent BSs on an infinite linear or circular cellular array. Variants of both models have been studied under various assumptions on the transmission schemes and the fading characteristics. 

In practical systems, perfect BS-cooperation or global processing is very difficult, if not impossible, to achieve. The main limitations are threefold: (i) limited backhaul capacity, (ii) local connectivity and (iii) imperfect CSI at the CS and the BSs.\footnote{Also the synchronization of the BSs as well as processing complexity and delay are limiting factors from an implementation perspective but are so far more or less neglected in the literature.} Therefore, most of the recent research targets the problem of constrained cooperation. For a detailed overview of this topic we refer to the surveys \cite{gesbert2010,Shamai2008}.
Information-theoretic implications of limited backhaul capacity have been studied separately for the uplink and downlink in \cite{Sanderovich2007b} and \cite{Simeone2009a}. Recently, the optimal amount of user data sharing between the BSs for the downlink with linear beamforming and backhaul constraints was studied in \cite{zakhour2010a}. 
The difficulties related to connecting a large number of BSs to a single CS have motivated the study of systems with only locally connected BSs \cite{Simeone2009,Levy2009,Levy2008a}. Several distributed algorithms for the uplink \cite{Aktas2008} and downlink \cite{Ng2008,Boccardi2008} have been proposed and it was shown that even with local BS connection near-optimal performance can be achieved with a reasonable amount of message passing and computational complexity. 

One of the most critical limitations of a practical network MIMO system, somehow overlooked compared to (i) and (ii), arises from the substantial overhead related to the acquisition of CSI (iii), indispensable to achieve the full diversity or multiplexing gains. This overhead becomes paramount, in particular for fast fading channels, when the number of antennas, sub-carriers, user terminals (UTs) or BSs grows \cite{Marzetta2009,Ramprashad2009,Ramprashad2009a,Caire2010}. Usually, CSI for the uplink is acquired through pilot signals sent by the UTs. This implies that a part of the coherence time of the channel needs to be sacrificed to obtain CSI with a sufficiently high quality. The inherent tradeoff between the resources dedicated to channel estimation and data transmission has been studied for the point-to-point MIMO channel \cite{hassibi03, Zheng2002} and the multi-user downlink \cite{kobayashi09a}. Recently, this problem was also addressed in the context of network MIMO systems, although with a different focus. In \cite{Caire2010,Ramprashad2009,Ramprashad2009a}, the authors compare several multi-cellular system architectures and conclude that the downlink performance of network MIMO systems is mainly limited by the inevitable acquisition of CSI (rather than by limited backhaul capacity). They also demonstrate that a conventional cellular system might outperform a network MIMO system under some circumstances assuming that the number of  \emph{coordinated} antennas and the used training overhead for both systems are the same. This means in essence that simply installing more antennas per BS can lead to higher performance improvements than installing costly backhaul infrastructure.

The imperfections detailed above call for robust strategies adapted to restricted BS-cooperation. Some schemes \cite{Kobayashi2009,Bjornson2009} rely on local CSI at the BSs and statistical CSI at the CS, whereas others \cite{Jing2008,Marsch2008b} consider serving only certain subsets of UTs with multiple BSs. Several BS-cooperation schemes have been studied in \cite{Marsch2009,Marsch2010} for the combination of limited backhaul capacity and imperfect CSI. The problem of ``pilot contamination'' caused by non-orthogonal training sequences in adjacent cells which can lead to significant inter-cell interference was addressed in \cite{Jose2009} and an optimized multi-cell precoding technique has been proposed. 

In this paper, we also consider limited BS-cooperation by focusing especially on the effects of imperfect CSI (iii). More precisely, we study the performance of the multi-cell uplink  with partially restricted cooperation assuming that:
\begin{itemize}
 \item The BSs act as oblivious relays which forward compressed versions of their received signals to the CS via orthogonal error- and delay-free backhaul links, each of fixed capacity $C\,\text{bits/channel use}$. 
\item The CS estimates the channel based on pilot tones sent by the UTs. 
\item The CS jointly processes the received signals from all BSs.
\end{itemize}
We consider a lower bound of the normalized ergodic mutual information of the network MIMO uplink channel with imperfect CSI and limited backhaul capacity, called the net ergodic achievable rate $R_\text{net}(\tau)$. For a given channel coherence time $T$, we attempt to find the optimal length $\tau^*$ of the pilot sequences for channel training which maximizes $R_\text{net}(\tau)$. As this optimization problem is in general intractable, we study a deterministic approximation $\overline{R}_\text{net}(\tau)$ of $R_\text{net}(\tau)$, based on large random matrix theory. 

The main contribution of this work is to show that optimizing $\overline{R}_\text{net}(\tau)$ instead of $R_\text{net}(\tau)$ is optimal in the large system limit. To this end, we provide a closed-form expression of the derivative of $\overline{R}_\text{net}(\tau)$ (Theorem~\ref{th:derivative}), prove the concavity of  $\overline{R}_\text{net}(\tau)$ for channel matrices with a doubly regular variance profile (Theorem~\ref{th:concavity}), and show that $\overline{\tau}^*$ which maximizes $\overline{R}_\text{net}(\tau)$ converges to $\tau^*$ in the large system limit (Theorem~\ref{th:convergence}). We further demonstrate by simulations that our asymptotic results yield tight approximations for systems of small dimensions with as little as three BSs and UTs.
In addition, we study the effects of limited backhaul capacity on the optimal channel training length. Since we assume that the CS estimates all channels based on the compressed observations from the BSs, the channel estimates are impaired by thermal noise \textit{and} quantization errors. Thus, increasing the backhaul capacity leads to improved channel estimates and, hence, smaller values of $\tau^*$.

The determination of the optimal training length $\tau^*$ in an uplink network MIMO setting with arbitrary path loss between the UTs and BSs and limited backhaul capacity appears to be a novel result, although we limit our investigation to a simple setting where $B$ cooperative BSs do not suffer from interference outside the network. The extension of this work to more realistic networks, such as clustered systems, is left to future investigations. Although the use of random matrix theory in the context of network MIMO is not new, see e.g. \cite{aktas06,huh10}, we present a novel application to an optimization problem in wireless communications.

The paper is structured as follows. The system model, including compression, channel training and data transmission, is described in Section~\ref{system_model}. The net ergodic achievable rate  $R_\text{net}(\tau)$ is defined in Section~\ref{net_rate} where we also present the deterministic approximation $\overline{R}_\text{net}(\tau)$ and discuss the optimization of the training length $\tau$. Numerical results and concluding remarks are given in Sections~\ref{numerical_results} and \ref{conclusion}, respectively.

\textit{Notations:} Boldface lowercase and uppercase letters designate column vectors and matrices, respectively. For a matrix $\Xm$,  $x_{ij}$ or $\LSB\Xm\RSB_{ij}$ denotes the $(i,j)$ entry of $\Xm$, $\left|\Xm \right|$ and $\trace\Xm$  denote the determinant and trace and $\Xm\tp$ and $\Xm\htp$ denote the transpose and complex conjugate transpose. For two matrices $\Xm$ and $\Ym$, $\Xm\otimes\Ym$ denotes the Kronecker (tensor) product. We denote an identity matrix of size $M$ as $\Id_M$ and  $\diag(x_1,\dots,x_M)$ is a diagonal matrix of size $M$ with the elements $x_i$ on its main diagonal. We use $\xv\sim\Cc\Nc\left(\mv,\Rm\right)$ to state that the vector $\xv$ has a circular symmetric complex Gaussian distribution with mean $\mv$ and covariance matrix $\Rm$. The natural logarithm is denoted by $\log(\cdot)$.
 
\section{System Model}\label{system_model}
\begin{figure}
\centering
\includegraphics[height=0.27\textheight]{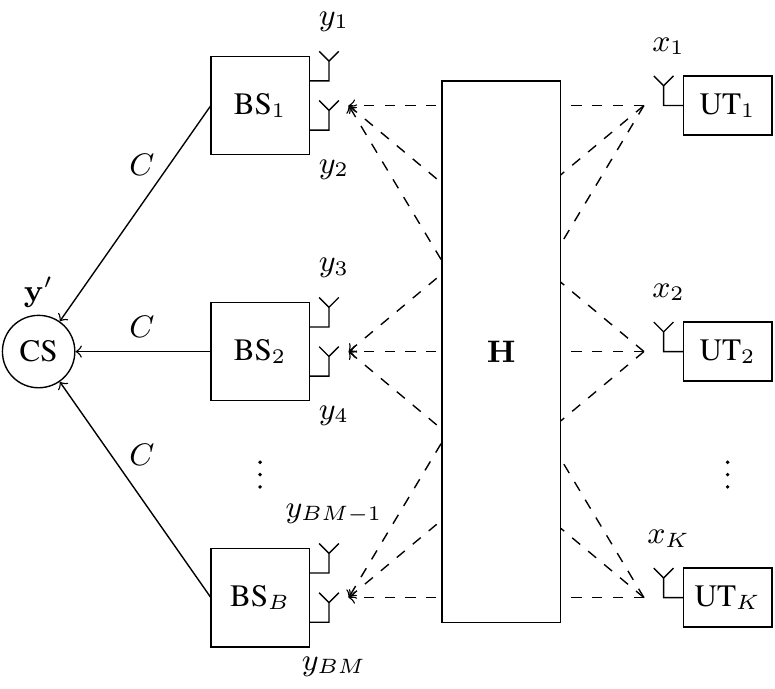}
\caption{Schematic system model for $M=2$ antennas per BS. The BSs compress and forward their received signals to the CS via orthogonal backhaul links of capacity $C\,\text{bits/channel use}$. The CS jointly processes the received data from all BSs.}
\label{fig:system_model}
\end{figure}

\subsection{Channel Model}
We consider a multi-cell frequency-selective fading uplink channel from $K$ single-antenna UTs to $B$ BSs with $M$ antennas each.\footnote{Our results can be easily extended to the case where each BS has a different number of antennas.} A schematic diagram of the channel model for $M=2$ is given in Fig.~\ref{fig:system_model}. Communication takes place simultaneously from all UTs to all BSs on $L$ parallel sub-carriers assuming an orthogonal frequency-division multiplexing (OFDM) transmission scheme. The stacked receive vector of all BSs on the $\ell$th sub-carrier $\yv(\ell)=\LSB y_1(\ell),\dots,y_{BM}(\ell)\RSB\tp\in\CC^{BM}$ at a given time reads
\begin{align}
 \yv(\ell) = \Hm(\ell)\xv(\ell) + \nv(\ell)
\end{align}
where $\xv(\ell)=\LSB x_1(\ell),\dots,x_K(\ell)\RSB\tp\in\CC^{K}$ is the vector of the transmitted signals of all UTs on sub-carrier $\ell$, $\nv(\ell)\sim\Cc\Nc\LB\zerov,\Id_{BM}\RB$ is a vector of additive noise and $\Hm(\ell)\in\CC^{BM\times K}$ is the aggregated channel matrix from all UTs to all BSs on the $\ell$th sub-carrier. 

We consider a discrete-time block-fading channel model where the channel remains constant for a coherence block of $T$ channel uses and then changes randomly from one block to the other. We let $T=T_c W_c$, where $W_c$ is the bandwidth per sub-carrier in Hz and $T_c$ the channel coherence time in seconds. Presuming that the bandwidth of each sub-carrier $W_c$ is on the order of the channel coherence bandwidth, that the antenna spacing at the BSs is sufficiently large and that the channels from the UTs to the BSs are uncorrelated, the channel matrices $\Hm_b(\ell)\in\CC^{M\times K}$, $b=1,\dots,B$, from the UTs to the BSs can be modeled as
\begin{align}\label{eqn:channel_model}
 \Hm_b(\ell) = \Wm_b(\ell)\,\diag\LB\sqrt{a_{b1}},\dots,\sqrt{a_{bK}}\RB
\end{align}
where $\Wm_b(l)\in\CC^{M\times K}$ is a standard complex Gaussian matrix and $a_{bk}$ denotes the inverse path loss between UT $k$ and BS $b$.\footnote{Note that the path loss is independent of the sub-carrier index $\ell$. This might not be the case for extremely large bandwidth but it is a reasonable assumption for most practical scenarios.} 
For later use, we define the matrix $\Vm\in\RR_+^{BM\times K}$ in the following way:
\begin{equation}\label{eqn:vprofile}
 \Vm = \Am\otimes\onev_M
\end{equation}
where $\Am\in\RR_+^{B\times K}$ is the inverse path loss matrix with elements $\{a_{bk}\}$ and $\onev_M$ is a $M$-dimensional column vector with all entries equal to one, such that the elements $\{v_{ij}\}$ of $\Vm$ satisfy $v_{ij}=a_{\lceil \frac Mi\rceil j}$. Under these assumptions, the elements $\{h_{ij}(\ell)\}$ of the matrix $\Hm(\ell)$ are independent circular symmetric complex Gaussian random variables with zero mean and variance $v_{ij}$, i.e., $h_{ij}(\ell)\sim\Cc\Nc(0,v_{ij})$. We refer to $\Vm$ as the \emph{variance profile} of the channel matrix $\Hm(l)$ and assume in the sequel that $\Vm$ is perfectly known at the CS while each BS $b$ only knows the distribution of its local channels $\Hm_b(\ell),\,\ell=1,\dots,L$. In a practical system, the channel coherence bandwidth might be significantly larger than the bandwidth of a sub-carrier so that $\{h_{ij}(\ell)\}$ would exhibit some correlation with respect to $\ell$. From a channel estimation perspective, the assumption of i.i.d.\@ channel coefficients represents a worst case since sub-carrier correlation cannot be exploited in the estimation process.

For simplicity, we assume Gaussian signaling with uniform power allocation, i.e., $x_k(\ell) \sim \Cc\Nc(0, P/L)$, i.i.d.\@ over $\ell$ and $k$, which is not necessarily optimal in the presence of channel estimation errors \cite{Yoo2006,hassibi03}. Although optimal power allocation over the sub-carriers would provide significant gains, it would require perfect channel knowledge at the UTs or some sort of feedback from the BSs/CS. Since we assume neither feedback nor CSI at the UTs and since the channel statistics are the same for all sub-carriers, uniform power allocation seems to be a reasonable choice. 

\subsection{Compression at the BSs}
The BSs are assumed to be oblivious to the applied codebooks of the UTs and forward compressed versions $y'_i(\ell)$ of their received signal sequences $y_i(\ell)$ to the CS via orthogonal backhaul links, each of capacity $C$ bits per channel use.\footnote{By \textit{orthogonal} backhaul links we mean here that there is no inter-backhaul interference. This is for example the case for a wired backhaul network with a dedicated link between the CS and each BS.} We also assume that the BSs and the CS have no prior knowledge of the instantaneous channel realizations. Under this setting, we consider a simple, sub-optimal compression scheme which neither exploits correlations between the received signals at different antennas nor adapts the employed quantization codebook to the actual channel realization. Thus, a single quantization codebook for the compression of each sequence $y_i(\ell)$ is used. This is in contrast to existing works, e.g. \cite{Sanderovich2009}, which rely on the assumption of full CSI at the BSs and the CS to apply optimized and channel dependent compression schemes. For a detailed discussion of different (distributed) compression schemes, we refer to \cite{Sanderovich2009,coso2009,Marsch2010} and references therein.

The rate-distortion function for the source $y_i(\ell)$ with squared error distortion is given as \cite[Theorem 10.2.1]{coverbook}
\begin{align}
 R_D\LB \sigma^2_i(\ell)\RB = \min_{\substack{f_{y'_i(\ell)|y_i(\ell)}:\\\expect{|y'_i(\ell)-y_i(\ell)|^2}\leq \sigma^2_i(\ell)}} I\LB y'_i(\ell);y_i(\ell)\RB
\end{align}
where the minimization is over all conditional probability density functions $f_{y'_i(\ell)|y_i(\ell)}$ satisfying the expected distortion constraint $\sigma^2_i(\ell)$. Similar to the so-called ``elementary compression scheme'' in \cite{Sanderovich2009}, our compression scheme is based on an underlying complex Gaussian ``test channel'' defined by 
\begin{align}\label{eqn:test_channel}
 y'_i(\ell) = y_i(\ell) + q_i(\ell)
\end{align}
where $q_i(\ell)\sim\Cc\Nc(0,\sigma^2_i(\ell))$. Note that the test channel \eqref{eqn:test_channel} used for the generation of the quantization codebooks is not optimal since the distribution of $y_i(\ell)=\sum_{j=1}^Kh_{ij}(\ell)x_j(\ell) +n_i(\ell)$ is not Gaussian. However, one can argue that in a large system with many UTs, the random variable $y_i(\ell)$ is almost Gaussian distributed and the performance degradation due to the sub-optimal choice of $f_{y'_i(\ell)|y_i(\ell)}$ is small. A simple upper bound of the rate distortion function is given by
\begin{align}\nonumber
 I(y'_i(\ell);y_i(\ell)) &= h(y'_i(\ell)) - h(y'_i(\ell)|y_i(\ell))\\\nonumber
&\leq \log\LB\pi e \LB\expect{|y_i(\ell)|^2}+\sigma^2_i(\ell)\RB\RB\\\nonumber &\qquad-\log\LB\pi e \sigma^2_i(\ell)\RB\\
&= \log \LB1 + \frac{1+\frac{P}{L}\sum_{j=1}^K v_{ij}}{\sigma^2_i(\ell)} \RB\label{eq:mutub}
\end{align}
where the inequality is obtained by upper-bounding the entropy of $y'_i(\ell)$ by the entropy of a complex Gaussian random variable with the same variance. We assume further that each BS uses $C/(ML)$ bits for the compression of each received complex symbol per antenna per sub-carrier. Replacing the left-hand side (LHS) of \eqref{eq:mutub} by $C/(ML)$, we can consequently overestimate the quantization noise variance $\sigma^2_i(\ell)$ by choosing
\begin{align}\label{eqn:quant_noise}
 \sigma^2_i = \sigma^2_i(\ell) = \frac{1+\frac{P}{L}\sum_{j=1}^K v_{ij}}{2^{\frac{C}{ML}}-1}.
\end{align}
Since the statistical distribution of $y_i(\ell)$ is the same for all sub-carriers, the quantization noise power $\sigma^2_i$ is also independent of $\ell$.  One can easily verify that the quantization noise vanishes for infinite backhaul capacity, i.e., $\sigma^2_i\to0$ for $C\to \infty$, and grows without bounds when the backhaul has zero capacity, i.e., $\sigma^2_i\to\infty$ for $C\to 0$.

We would like to point out that the field of distributed compression with imperfect CSI is to the best of our knowledge a largely unexplored area. It is for example not clear if each BS should estimate its local channels and forward compressed versions of its estimates to the CS or if the CS should estimate all channels based on compressed signals from the BSs, as assumed in this work.

\subsection{Channel Training}
Similar to \cite{hassibi03}, each channel coherence block of length $T$ is split into a phase for channel training and a phase for data transmission. During the training phase of length $\tau$, all $K$ UTs broadcast orthogonal sequences of known pilot symbols of equal power $P/L$ on all sub-carriers. The orthogonality of the training sequences imposes $\tau\geq K$. We assume that the CS estimates the channels $h_{ij}(\ell)$ from all UTs to all BSs based on the observations
\begin{equation}\label{eqn:estimation}
 r_{ij}(\ell) = \sqrt{\frac{\tau P}{L}}h_{ij}(\ell) + s_{ij}(\ell)
\end{equation}
where $s_{ij}(\ell)\sim\Cc\Nc(0,1+\sigma_i^2)$ captures the effects of the thermal noise at the BS-antennas and the quantization error on the backhaul links. For details on how the scalar estimation channel \eqref{eqn:estimation} is obtained, we refer the reader to \cite{hassibi03}. It becomes clear from the last equation that the quantization noise degrades the channel estimate. Thus, the backhaul capacity $C$ has a significant influence on the optimal training length $\tau^*$. This point will be further discussed in Section~\ref{numerical_results}. Computing the minimum mean square error (MMSE) estimate of $h_{ij}(\ell)$ given the observation $r_{ij}(\ell)$, we can decompose $h_{ij}(\ell)$ into the estimate $\hat{h}_{ij}(\ell)$ and the independent estimation error $\tilde{h}_{ij}(\ell)$, such that
\begin{equation}
 h_{ij}(\ell) = \hat{h}_{ij}(\ell) + \tilde{h}_{ij}(\ell).
\end{equation}
The variance of the estimated channel $\hat{v}_{ij}(\tau)$ and the variance of the estimation error $\tilde{v}_{ij}(\tau)$ are respectively given as
\begin{align}\label{eqn:est_var}
 \hat{v}_{ij}(\tau)&\defines\expect{|\hat{h}_{ij}(\ell)|^2}=\frac{\tau \frac{P}{L} v_{ij}^2}{\tau \frac{P}{L} v_{ij}+1+\sigma_i^2}\qquad\forall \ell\\\label{eqn:err_var}
 \tilde{v}_{ij}(\tau)&\defines\expect{|\tilde{h}_{ij}(\ell)|^2}=\frac{v_{ij}(1+\sigma_i^2)}{\tau \frac{P}{L} v_{ij}+1+\sigma_i^2}\qquad\forall \ell.
\end{align}
Denote $\hat{\Vm}(\tau)$ and $\tilde{\Vm}(\tau)$ the variance profiles of the estimated channel $\hat{\Hm}(\ell)$ and the estimation error $\tilde{\Hm}(\ell)$, respectively. One can easily verify that the total energy of the channel is conserved since 
\begin{equation}
 \Vm = \hat{\Vm}(\tau) + \tilde{\Vm}(\tau)\ .
\end{equation} 

\subsection{Data Transmission}\label{sec:datatx}
 In each channel coherence block, the UTs broadcast their data simultaneously during $T-\tau$ channel uses. The CS jointly decodes the messages from all UTs, leveraging the previously computed channel estimate  $\hat{\Hm}(\ell)$. With the knowledge of $\hat{\Hm}(\ell)$, the CS ``sees'' in its received signal $\yv'(\ell)=\LSB y'_1(\ell),\dots,y'_{BM}(\ell)\RSB\tp$ the useful term $\hat{\Hm}(\ell)\xv(l)$ and the overall noise term $\zv(\ell) = \tilde{\Hm}(\ell)\xv(\ell) + \nv(\ell) + \qv(\ell)$, i.e.,
\begin{align}\label{eqn:effective_channel}
 \yv'(\ell) = \hat{\Hm}(\ell)\xv(\ell) + \zv(\ell)
\end{align}
where the quantization noise vector $\qv=\LSB q_1(\ell),\dots,q_{BM}(\ell)\RSB\tp$ is defined by \eqref{eqn:test_channel}. Since the statistical distributions of all sub-carriers, signals and noise are i.i.d. with respect to the index $\ell$, we will hereafter omit the dependence on $\ell$ and consider a single isolated sub-carrier.

\section{Net Ergodic Achievable Rate}\label{net_rate}
The capacity of the channel \eqref{eqn:effective_channel} is not explicitly known. We consider therefore a lower bound of the normalized ergodic mutual information $\frac{1}{BM}I\LB\yv';\xv|\hat{\Hm}\RB$, referred to hereafter as the \emph{ergodic achievable rate} $R(\tau)$. This lower bound is in essence obtained by overestimating the detrimental effect of the estimation error, treating the total noise term $\zv$ as independent complex Gaussian noise with covariance matrix $\Km_z(\tau)\in\RR_+^{BM\times BM}$, given as
\begin{align}\label{eqn:covariance}\nonumber
\Km_z(\tau) &= \expect{\zv\zv\htp}\\ &= \diag\LB1+\sigma^2_i + \frac{P}{L}\sum_{j=1}^K \tilde{v}_{ij}(\tau)\RB_{i=1}^{BM}.
\end{align}
Thus, the ergodic achievable rate can be written as \cite{Yoo2006,hassibi03}
\begin{align}\label{eqn:ergodic_achievable_rate}
R(\tau) = \frac{1}{BM}\expectl{\hat{\Hm}}{\log\left|\Id_{BM} + \frac{P}{L}\overline{\Hm}(\tau)\overline{\Hm}(\tau)\htp \right|}
\end{align}
where we have defined the effective channel $\overline{\Hm}(\tau)$ as
\begin{align}\label{eq:eff_chn}
 \overline{\Hm}(\tau) = \Km_z^{-\frac12}(\tau)\hat{\Hm} .
\end{align}
 
Note that the ergodic achievable rate does not account for the fact that only a fraction $(1-\tau/T)$ of the total coherence block length can be used for data transmission. Our goal is thus to find the optimal training length $\tau^*$, maximizing the \emph{net ergodic achievable rate}
\begin{align}
 R_\text{net}(\tau) \ \defines\ \left(1-\frac{\tau}{T}\right) R(\tau).
\end{align}
Here, the difficulty consists in computing the ergodic achievable rate $R(\tau)$ explicitly. Since a closed-form expression of $R(\tau)$ for finite dimensions of the channel matrix $\Hm$ seems intractable, we resort to an approximation based on the theory of large random matrices. We will demonstrate shortly that this approximation, although only asymptotically tight, yields very close approximations for even small values of $B$, $M$, $K$ and $L$.

\subsection{Deterministic Equivalent}
In this section, we present a deterministic equivalent approximation $\overline{R}(\tau)$ of $R(\tau)$ in the large system limit, i.e., for $K,BM,L\to\infty$ at the same speed. Denote $N=BM$ the product of the number of BSs and the number of antennas per BS. The notation $K\to\infty$ will refer in the sequel to the following two conditions on $K,N$ and $L$:
\begin{align}\label{eq:pace}\nonumber
 &0 < \liminf_{K\to\infty} \frac {N}{K}\leq \limsup_{K\to\infty} \frac {N}{K}  <  \infty\ \\
 &0 < \liminf_{K\to\infty} \frac LK \leq  \limsup_{K\to\infty}  \frac LK  <  \infty.
\end{align}
Define $\overline{\Vm}(\tau) = \Km_z^{-1}(\tau)\hat{\Vm}(\tau)$ the variance profile of the effective channel $\overline{\Hm}(\tau)$ with elements
\begin{align}\label{eqn:veff}
 \overline{v}_{ij}(\tau) = \frac{\hat{v}_{ij}(\tau)}{1+\sigma_i^2 + \frac PL\sum_{\ell=1}^K\tilde{v}_{i\ell}(\tau)}
\end{align}
and consider the following $N\times N$ matrices
\begin{align}
 \Dm_j(\tau) = \diag\LB \overline{v}_{1j}(\tau),\dots,\overline{v}_{Nj}(\tau) \RB,\quad j=1,\dots,K.
\end{align}
Denote by $\CC_+=\{z\in\CC:\Im(z)>0\}$, and by ${\mathcal S}$ the class of functions $f$ analytic over $\CC\setminus\RR_+$, such that for $z\in\CC_+$, $f(z)\in\CC_+$ and $zf(z)\in\CC_+$,  and $\lim_{y\rightarrow \infty} -\mathbf{i}y f(\mathbf{i} y)=1$, where $\mathbf{i}=\sqrt{-1}$.\footnote{Such functions are known to be Stieltjes transforms of probability measures over $\mathbb{R}_+$ - see for instance \cite[Proposition 2.2]{hachem07}.}
We are now in position to state the deterministic approximation $\overline{R}(\tau)$ of $R(\tau)$ based on a direct application of \cite[Theorem 2.3]{hachem08} (see also \cite[Theorems 2.4 and 4.1]{hachem07}) to our channel model.

\begin{theorem}[Deterministic Equivalent]\label{thm:det_equ}
Let $\tau>0$. Assume that $K$, $N$ and $L$ satisfy \eqref{eq:pace} and $0\leq \overline{v}_{ij}(\tau)<v_\text{max}<\infty\,\forall i,j$. Then:
\begin{enumerate}
\item[(i)] The following implicit equation: 
\begin{align}\label{eq:T}
\Tm(z) = \LB \frac{1}{K}\sum_{j=1}^K\frac{\Dm_j(\tau)}{1+\frac 1K \trace \Dm_j(\tau) \Tm(z) }-z \Id_{N} \RB^{-1}
\end{align}
admits a unique solution $\Tm(z)=\diag\LB t_1(z),\dots,t_N(z)\RB$ such that $(t_1(z),\dots,t_N(z))\in {\mathcal S}^N$. 
\item[(ii)] Let $P>0$. Denote $\Tm_P=\Tm(-\frac{L}{KP})$ and consider the quantity:
\begin{align}\nonumber
 \overline{R}(\tau) =  & \frac{1}{N}\sum_{j=1}^K\log\LB1+\frac 1K \trace \Dm_j(\tau) \Tm_P\RB\\\nonumber &\ -\frac{1}{N}\log\det\LB\frac{L}{KP}\Tm_P\RB\\\label{eq:rdet}
 &\ - \frac{1}{N}\sum_{j=1}^K\frac{\frac 1K \trace \Dm_j(\tau) \Tm_P}{1+\frac 1K \trace \Dm_j(\tau) \Tm_P}.
\end{align}
Then, the following holds true:
\begin{align}
 R(\tau) - \overline{R}(\tau)\xrightarrow[K\rightarrow \infty]{} 0.
\end{align}
\end{enumerate}
\end{theorem}

\subsection{Optimization of the training length $\tau$}
In this section, we consider the optimization of the training length $\tau$ with the goal of maximizing the net ergodic achievable rate $R_\text{net}(\tau)$. In order to find the optimal training length $\tau^*$ for a given coherence block length $T$, we wish to solve the following optimization problem:
\begin{align}
&\text{maximize} \qquad R_\text{net}(\tau)\\\nonumber
&\text{subject to}\qquad  K\leq\tau\leq T.
\end{align}

As this optimization problem is intractable for finite dimensions, we pursue the following approach:
\begin{enumerate}
 \item We find $\overline{\tau}^*$ maximizing the deterministic approximation $\overline{R}_\text{net}(\tau)=\LB1-\frac\tau T\RB \overline{R}(\tau)$.
\item We show that $R_\text{net}(\tau^*)-\overline{R}_\text{net}(\overline{\tau}^*)\to 0$ and $\tau^*-\overline{\tau}^*\to 0$ as $K\to\infty$.
\item We verify by simulations that $\overline{\tau}^*$ is very close to $\tau^*$ for even small values of $K,N$ and $L$.
\end{enumerate}

We start by establishing the concavity of $\overline{R}_\text{net}(\tau)$, our new objective function.
Denote\footnote{We use $f'(x)$ to denote the first derivative of the function $f(x)$, i.e.,  $f'(x)=\frac{d\,f(x)}{d\, x}$.}
\begin{align}\label{eq:vareffd}\nonumber
 &\overline{v}'_{ij}(\tau) =\\ &\ \frac{\hat{v}'_{ij}(\tau)\LSB1+\sigma_i^2 + \frac PL\sum_{j=1}^K\tilde{v}_{ij}(\tau)\RSB-\hat{v}_{ij}(\tau)\frac{P}{L}\sum_{j=1}^K\tilde{v}'_{ij}(\tau)}{\LSB1+\sigma_i^2 + \frac PL\sum_{j=1}^K\tilde{v}_{ij}(\tau)\RSB^2}
\end{align}
where
\begin{align}\label{eq:vard}
 \hat{v}'_{ij}(\tau) = -\tilde{v}'_{ij}(\tau) = \frac{\frac PLv_{ij}^2\LB1+\sigma_i^2\RB}{\LB1+\sigma^2_i+\tau\frac PL v_{ij}\RB^2}
\end{align}
and define the matrices
\begin{align}
 \Dm_j'(\tau) = \diag\LB\overline{v}'_{1j}(\tau),\dots,\overline{v}'_{Nj}(\tau)\RB,\quad j=1,\dots,K. 
\end{align}
A simple composition rule \cite[Exercise 3.32 (b)]{boyd_cvx} states that the product of a positive decreasing linear function and a positive increasing concave function is also concave. In order to prove the concavity of $\overline{R}_\text{net}(\tau)=(1-\frac\tau T)\overline{R}(\tau)$, it is thus sufficient to show that $\overline{R}(\tau)$ is an increasing concave function in $\tau$. A sufficient condition for concavity is $\overline{R}''(\tau)\leq 0$. We begin by considering the first derivative $\overline{R}'(\tau)$, which allows for a simple concise closed-from expression as provided by the next theorem:

\begin{theorem}[Derivative]\label{th:derivative} Under the same conditions as for Theorem~\ref{thm:det_equ}, the first derivative of $\overline{R}(\tau)$ permits the explicit expression
 \begin{align}
  \overline{R}'(\tau) = \frac1N\sum_{j=1}^K\frac{\frac1K \trace\Dm'_j(\tau)\Tm_P}{1+\frac1K\trace\Dm_j(\tau)\Tm_P}
\end{align}
where $\Tm_P=\Tm(-\frac{L}{KP})$ is given by Theorem~\ref{thm:det_equ}~(i). Moreover, for any $P,\tau > 0$, $\overline{R}(\tau)$ is an increasing function, i.e., 
\begin{align}
 \overline{R}'(\tau) > 0.
\end{align}
\end{theorem}
\begin{IEEEproof}
See Appendix~\ref{th:derivative_proof}.
\end{IEEEproof}

Despite the simplicity of the expression of $\overline{R}'(\tau)$ in Theorem~\ref{th:derivative}, it seems intractable to show that $\overline{R}_\text{net}''(\tau)\leq 0$ for channel matrices with a general variance profile. This is due to the fact that not only $\Dm_j(\tau)$ depends on $\tau$, but also $\Tm_P$. The matrix $\Tm_P$ is in general given as the solution of an implicit equation which can only be determined numerically, e.g. by a fixed-point algorithm. It is thus difficult to infer the behavior of $\Tm_P$ with respect to $\tau$. However, one can show for the particular case of a doubly regular variance profile that $\overline{R}(\tau)$ is indeed concave.

\begin{theorem}[Concavity]\label{th:concavity}
 Let $P,\tau>0$. Assume that $N=K$ and that $\overline{\Vm}(\tau)$ is a doubly regular matrix which satisfies the following regularity condition:
\begin{align}
\Kc(\tau) = \frac1N\sum_{i=1}^N\overline{v}_{ik}(\tau) = \frac1N\sum_{j=1}^N\overline{v}_{\ell j}(\tau)\quad\forall k,\ell.
\end{align}
Then, $\overline{R}(\tau)$ is a strictly concave function.
\end{theorem}
\begin{IEEEproof}
 See Appendix~\ref{th:concavity_proof}.
\end{IEEEproof}
\begin{remark}\label{claim}
Based on our simulation results, we conjecture that Theorem~\ref{th:concavity} also holds for non doubly regular variance profiles $\overline{\Vm}(\tau)$. Intuitively, $\overline{R}(\tau)$ being a concave function means nothing else than that channel training shows diminishing returns. That is, the marginal benefit of each training symbol decreases until the channel estimation becomes nearly perfect. The previous argument can be made clear considering the two extreme cases $\tau=0$ and $\tau\to\infty$. One can easily verify that $\Dm_j(0)=\zerov$ while $\Dm_j'(0)>\zerov$. This implies $\overline{R}'(0)>0$, i.e., channel training increases the ergodic achievable rate. On the other hand, as $\tau\to\infty$, $\Dm_j'(\tau)\to\zerov$, so that also $\overline{R}'(\tau)\to 0$, i.e., the marginal benefit of channel training vanishes. It is thus justified to conjecture that $\overline{R}'(\tau)$ is a decreasing function of $\tau$ and hence $\overline{R}(\tau)$ a concave function.
\end{remark}

As a consequence of Theorem~\ref{th:concavity} and Remark~\ref{claim}, we assume that $\overline{R}_\text{net}(\tau)$ takes its global maximum in $(0,T]$ and the optimal training length $\overline{\tau}^*$ can be determined as the solution of
\begin{align}     
 \overline{R}_\text{net}'(\tau) = \LB1-\frac\tau T\RB \overline{R}'(\tau) - \frac1T\overline{R}(\tau)=0.
\end{align}
The value $\overline{\tau}^*$ can now be easily found, e.g. via the bisection method. It remains to show that the optimal training length $\overline{\tau}^*$ which maximizes $\overline{R}_\text{net}(\tau)$ is asymptotically optimal for the original objective function $R_\text{net}(\tau)$. This is done in the next theorem.

\begin{theorem}[Convergence]\label{th:convergence}
Let $\tau^* = \arg\max_{\tau\in[0,T]} R_\text{net}(\tau)$ and  $\overline{\tau}^* = \arg\max_{\tau\in[0,T]} \overline{R}_\text{net}(\tau)$. Then, under the same conditions as for Theorem~\ref{thm:det_equ}, the following holds true:
{\begin{itemize}
 \item[(i)]\begin{align}R_\text{net}(\tau^*)-\overline{R}_\text{net}(\overline{\tau}^*)\xrightarrow[K\rightarrow \infty]{} 0.\end{align}
\item[(ii)] Further assume that $\overline{\Vm}(\tau)$ is a doubly regular matrix which satisfies the conditions of Theorem~\ref{th:concavity}. Then,  \begin{align}\tau^*-\overline{\tau}^*\xrightarrow[K\rightarrow \infty]{} 0\end{align}
where $\overline{\tau}^*$ is given as the solution to \begin{align}\overline{R}_\text{net}'(\tau) = \LB1-\frac\tau T\RB \overline{R}'(\tau) - \frac1T\overline{R}(\tau)=0 \end{align}
with $\overline{R}(\tau)$ and $\overline{R}'(\tau)$ given by Theorem~\ref{thm:det_equ} (ii) and Theorem~\ref{th:derivative}, respectively.
\end{itemize}}
\end{theorem}
\begin{IEEEproof}
See Appendix~\ref{th:convergence_proof}.
\end{IEEEproof}

Theorem~\ref{th:convergence}~(i) merely states that the maximum point of $R_\text{net}(\tau)$ can be arbitrarily closely approximated by the maximum point of 
$\overline{R}_\text{net}(\tau)$. This result is independent of the structure of the variance profile $\overline{\Vm}(\tau)$. Theorem~\ref{th:convergence}~(ii) provides a simple way to compute $\overline{\tau}^*$ and states that this value is also asymptotically optimal for $\overline{R}_\text{net}(\tau)$. However, this result requires $\overline{\Vm}(\tau)$ to be a doubly regular matrix. Both results together imply that optimizing $\overline{R}_\text{net}(\tau)$ is asymptotically identical to optimizing $\overline{R}_\text{net}(\tau)$. We show in the next section via simulations that Theorem~\ref{th:concavity} and Theorem~\ref{th:convergence} also hold for non doubly regular variance profiles. 

\section{Numerical Results}\label{numerical_results}
\begin{figure}
\begin{center}
 \includegraphics[height=0.2\textheight]{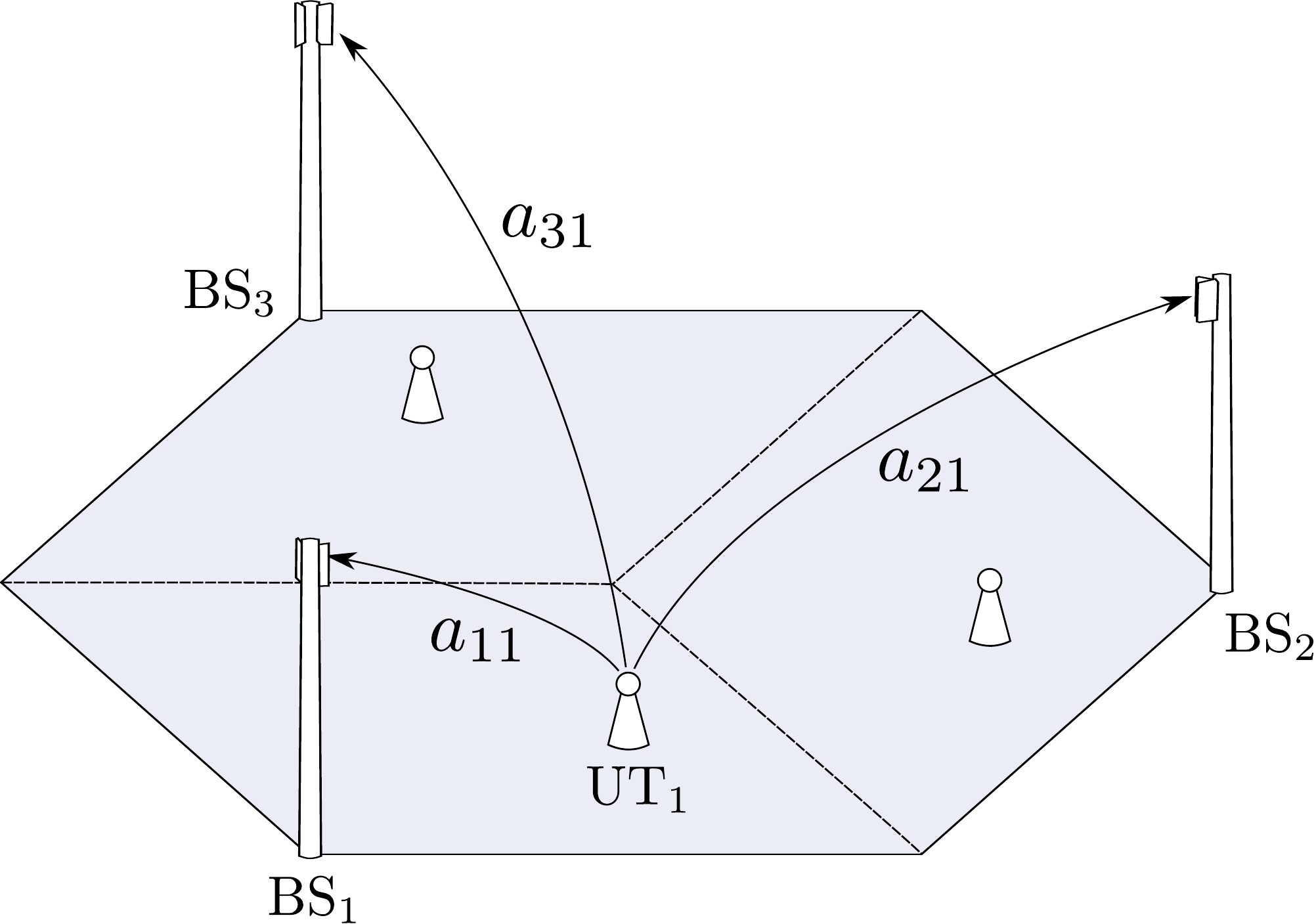} 
\end{center}
\caption{Cellular example with $B=3$ BSs and $K=3$ UTs.\label{fig:cell}}
\end{figure}

In order to show the validity of our analysis in the preceding sections, we consider a simple cellular system consisting of $B=3$ BSs with $M=2$ antennas and $K=3$ UTs, as shown in Fig.~\ref{fig:cell}. The locations of the UTs are randomly chosen according to a uniform distribution. The inverse path loss factor $a_{bk}$ between UT $k$ and BS $b$ is given as $a_{bk} = d_{bk}^{-3.6}$, where $d_{bk}$ is the distance between UT $k$ and BS $b$, normalized to the maximum distance within a cell. We consider one random snapshot of user distributions, resulting in the inverse path loss matrix
\begin{align}
 \Am = \begin{pmatrix}
        2.9775 &    0.0385&    1.6055\\
        0.2512&    2.7826&    0.1759\\
        0.0615&    0.0492&    1.6376
       \end{pmatrix}.
\end{align}
In the sequel, we assume $\Am$ fixed while we average over many independent realizations of the channel matrix $\Hm$. The cell edge signal-to-noise-ratio is defined as $\text{SNR}=\expect{|x_i(\ell)|^2}/\expect{|n_i(\ell)|^2}=P/L$. Unless otherwise stated, we assume $T=1000$  and $L=1$.

\begin{figure}
\centering
\includegraphics[height=0.25\textheight]{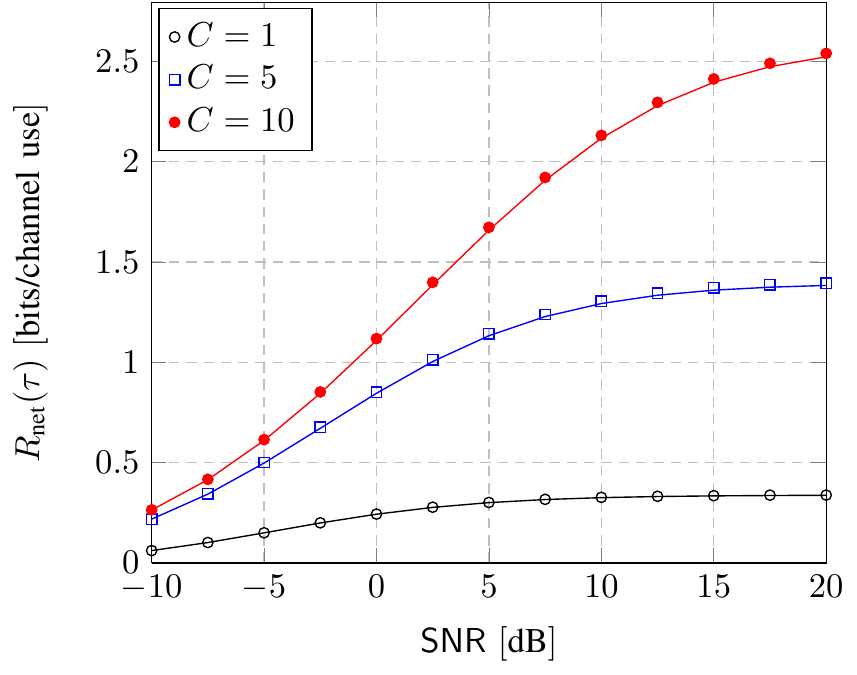}
\caption{Net ergodic achievable rate $R_\text{net}(\tau)$ vs SNR for $\tau=40$ and $T=1000$. The markers are obtained by simulations, the solid lines correspond to the deterministic equivalent $\overline{R}_\text{net}(\tau)$.\label{fig:Rnet_vs_snr}}
\end{figure}

Fig.~\ref{fig:Rnet_vs_snr} depicts the net ergodic achievable rate $R_\text{net}(\tau)$ and its deterministic equivalent approximation $\overline{R}_\text{net}(\tau)$ by Theorem~\ref{thm:det_equ}~(ii) as a function of the SNR for a fixed training length of $\tau=40$ and different values of the backhaul capacity $C=\{1,5,10\}\,\text{bits/channel use}$. Clearly, $\overline{R}_\text{net}(\tau)$ gives a very tight approximation of $R_\text{net}(\tau)$ over the full range of SNR. The effect of limited backhaul is particularly visible at high SNR where all curves saturate. 

\begin{figure}
\centering
\includegraphics[height=0.25\textheight]{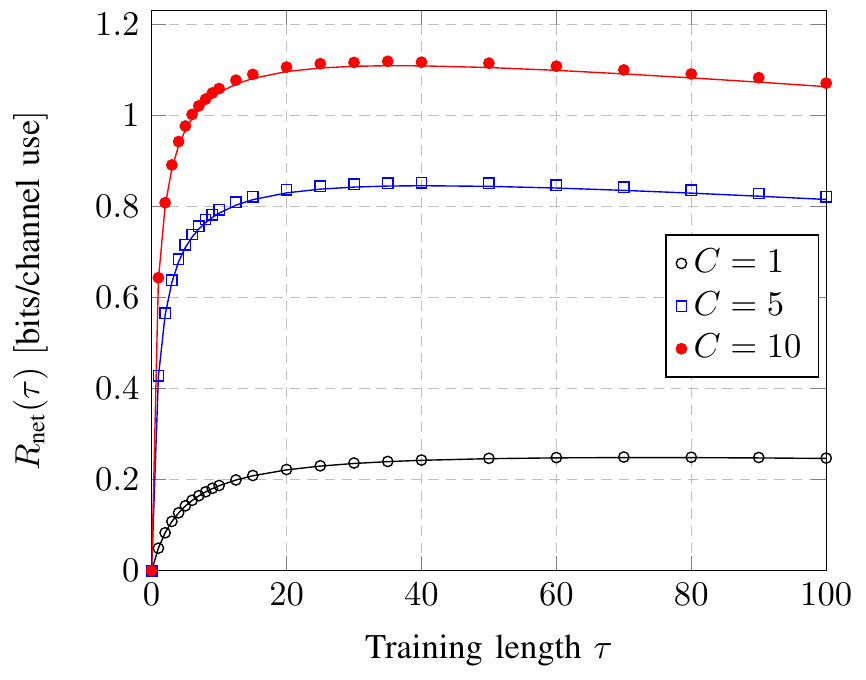}
\caption{Net ergodic achievable rate $R_\text{net}(\tau)$ vs training length $\tau$ for $\text{SNR}=0\,\text{dB}$ and $T=1000$. The markers are obtained by simulations, the solid lines correspond to the deterministic equivalent $\overline{R}_\text{net}(\tau)$.\label{fig:Rnet_vs_t}}
\end{figure}

\begin{figure}
\centering
\includegraphics[height=0.25\textheight]{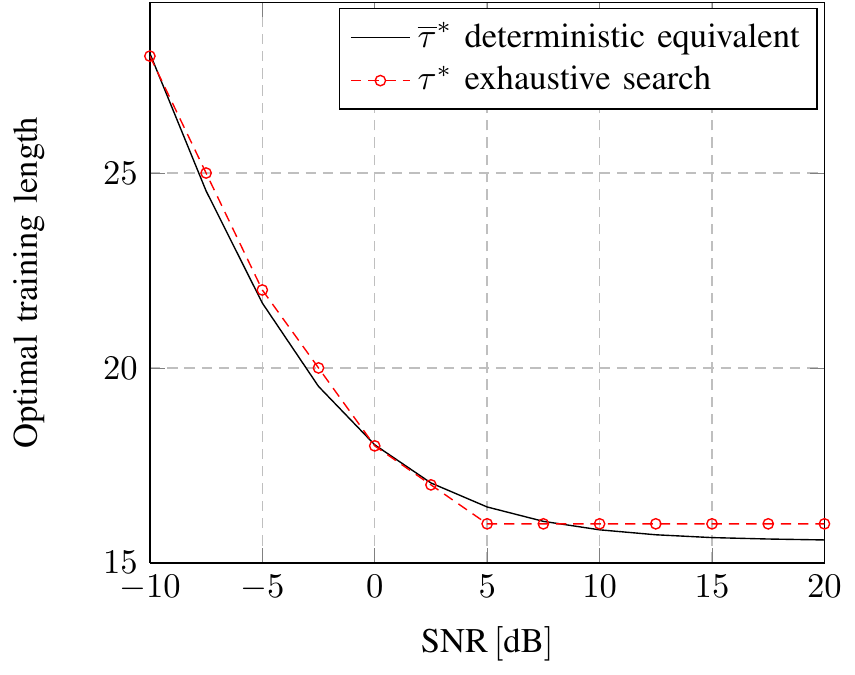}
\caption{Optimal training length $\tau^*$ and $\overline{\tau}^*$ vs SNR for $C=1\,\text{bits/channel use}$ and $T=100$. The solid line corresponds to $\overline{\tau}^*$ maximizing $\overline{R}_\text{net}(\tau)$, the dashed line corresponds to $\tau^*$ maximizing $R_\text{net}(\tau)$ and is obtained by an exhaustive search based on Monte Carlo simulations.\label{fig:topt_vs_snr}}
\end{figure}

For the same set of parameters and $\text{SNR}=0\,\text{dB}$, we show in Fig.~\ref{fig:Rnet_vs_t}  $R_\text{net}(\tau)$ and  $\overline{R}_\text{net}(\tau)$ as a function of the training length $\tau$.  This plot validates Theorem~\ref{th:concavity} and the corresponding remark as $\overline{R}_\text{net}(\tau)$ is obviously a concave function. Moreover, since the curves of $\overline{R}_\text{net}(\tau)$  and $R_\text{net}(\tau)$ match very closely, it is reasonable to assume that both take a similar maximum value at a similar value of $\tau$. 
The validity of Theorem~\ref{th:convergence} is demonstrated in Fig.~\ref{fig:topt_vs_snr} which shows the optimal training length $\tau^*$, found by an exhaustive search based on Monte Carlo simulations, and the training length $\overline{\tau}^*$ which maximizes $\overline{R}_\text{net}(\tau)$ as a function of the SNR for $C=1\,\text{bits/channel use}$ and $T=100$. The differences between both values, although very small, are mainly due to the exhaustive search over a necessarily discrete set of values of $\tau$.

\begin{figure}
\centering
\includegraphics[height=0.25\textheight]{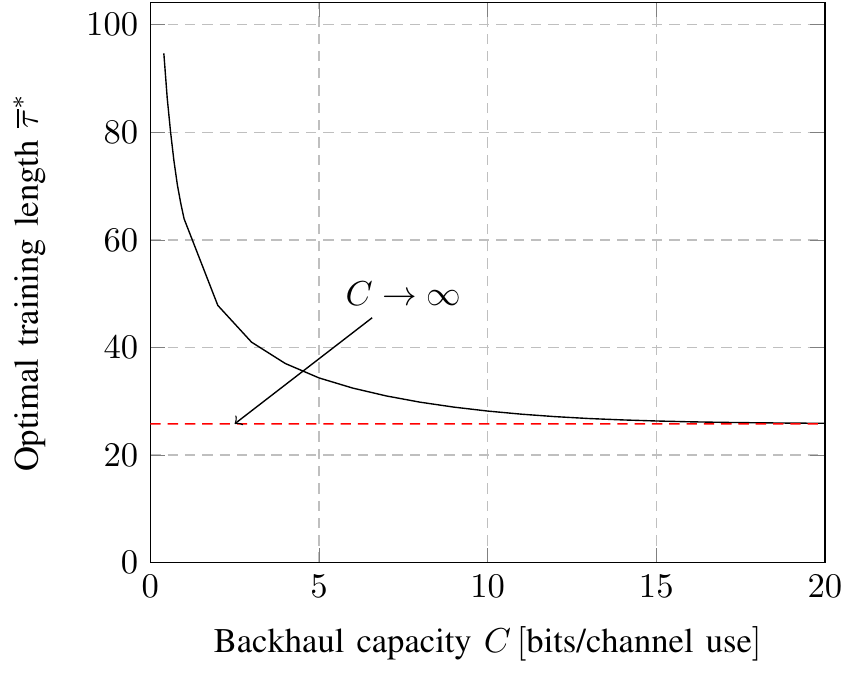}
\caption{Optimal training length $\overline{\tau}^*$ vs backhaul capacity $C$ for $\text{SNR}=10\, \text{dB}$ and $T=1000$. \label{fig:topt_vs_C}}
\end{figure}

\begin{figure}
\centering
\includegraphics[height=0.25\textheight]{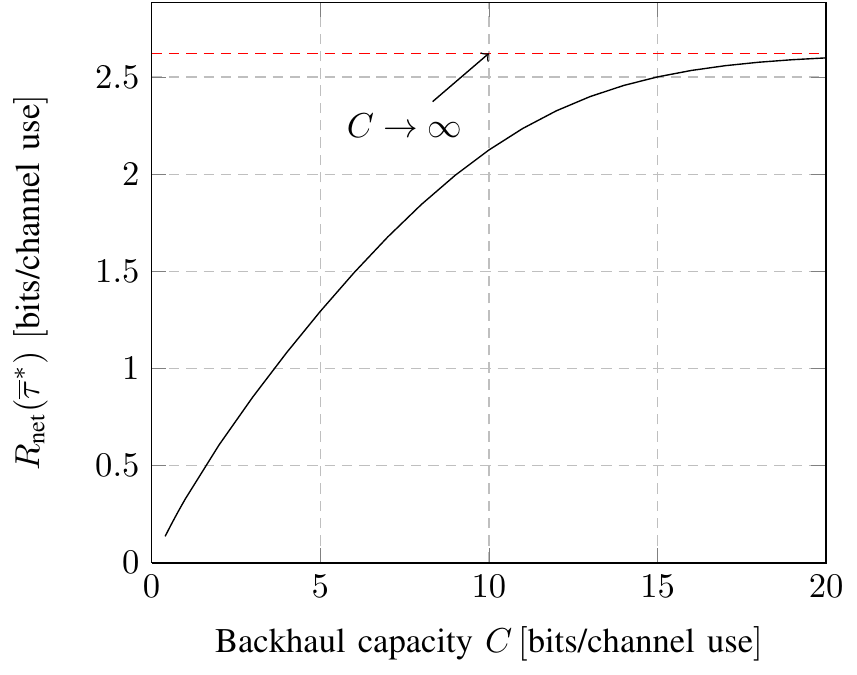}
\caption{Net ergodic achievable rate $R_\text{net}(\overline{\tau}^*)$ with optimal channel training $\overline{\tau}^*$ vs backhaul capacity $C$ for $\text{SNR}=10\, \text{dB}$ and  $T=1000$. \label{fig:Rnet_vs_C}}
\end{figure}

Fig.~\ref{fig:topt_vs_C} shows the dependence of the optimal training length $\overline{\tau}^*$ on the backhaul capacity $C$ for a fixed $\text{SNR}=10\,\text{dB}$. One can see that $\overline{\tau}^*$ is a decreasing function of $C$ which converges quickly to particular value corresponding to infinite capacity backhaul links. The reason for this is the following. The CS estimates the channel coefficients based on the quantized training signals received by the BSs. The channel estimate is hence impaired by thermal noise and quantization errors. Therefore, increasing $C$ results in better channel estimates and reduces the necessary training length. For infinite backhaul capacity, the optimal training length is only dependent on the SNR.
In a similar flavor, Fig.~\ref{fig:Rnet_vs_C} depicts $R_\text{net}(\overline{\tau}^*)$ as a function of the backhaul capacity $C$. We notice the inefficient utilization of the backhaul links due to sub-optimal compression since the net ergodic achievable rate per BS, i.e., $M\times R_\text{net}(\overline{\tau}^*)$, is much lower than the necessary backhaul capacity. For example, it takes $C=20\,\text{bits/channel use}$ of backhaul capacity to achieve a rate per BS of $2\times R_\text{net}(\overline{\tau}^*)\approx 5.2 \,\text{bits/channel use}$.

\section{Conclusion}\label{conclusion}
In this work, we have considered a frequency-selective fading network MIMO uplink channel with arbitrary path losses between the UTs and BSs and finite capacity backhaul links. Using a close approximation of the net ergodic achievable rate based on random matrix theory, we have studied the optimal tradeoff between the resources used for channel training and data transmission. Although the asymptotic results are proved to be tight only in the large system limit,  our numerical examples show that they provide close approximations even for small system dimensions. Our results also show that limited backhaul capacity has a significant impact on the optimal training length. We wish to conclude the paper by pointing out some shortcomings of our system model which remain as future investigations.

\subsubsection{Backhaul links and cooperation}
 A relevant question is how a BS should decide whether to cooperate by forwarding its received data to some central processor or to process its received signals alone. In our model, the net throughput vanishes with a decreasing backhaul capacity although each BSs could theoretically decode a part of the received messages alone. Future work, also motivated by the recent results in \cite{Marsch2010,Marsch2008d}, comprises the investigation of flexible schemes which adapt the degree of cooperation according to some statistical side-information about the channels, backhaul limitations, quality of CSI, etc. 
 
\subsubsection{Inter-cluster interference}
We have considered a multi-cell network composed of $B$ cooperative cells without inter-cell interference. In a real system, also the effects of non-orthogonal training sequences leading to ``pilot contamination'' \cite{Jose2009,Marzetta2009} constitute an important issue for practical system design. Both aspects need to be taken into account for a more realistic performance evaluation of network MIMO systems.

\appendices
\section{Proof of Theorem~\ref{th:derivative}}\label{th:derivative_proof}
We start by defining the following auxiliary variables $\delta_j = \frac1K\trace\Dm_j(\tau)\Tm_P,\ j=1,\dots,K$. Using this definition, we can re-write $\overline{R}(\tau)$ in \eqref{eq:rdet} as
\begin{align}\nonumber
 \overline{R}(\tau) =& \frac1N \sum_{j=1}^K\LSB\log(1+\delta_j) - \frac{\delta_j}{1+\delta_j}\RSB \\ &\ -\frac1N\log\det\LB\frac{L}{KP}\Tm_P\RB.
\end{align}
We define $\delta'_j = \frac{d\, \delta_j}{d\, \tau} =  \frac1K\trace\Dm'_j(\tau)\Tm_P + \frac1K\trace\Dm_j(\tau)\Tm_P'$, where $\Tm_P' = \frac{d\, }{d\, \tau}\Tm_P$. Taking the derivative of $\overline{R}(\tau)$ with respect to $\tau$ yields
\begin{align}\label{eq:intermediate}
 \overline{R}'(\tau) = \frac1N \sum_{j=1}^K\LSB\frac{\delta_j\delta_j'}{(1+\delta_j)^2}\RSB-\frac1N\trace\Tm_P^{-1}\Tm_P'.
\end{align}
This expression can be further simplified by re-writing the definition of $\Tm_P$ as a function of $\delta_j$:
\begin{align}
 \Tm_P = \LB\frac{L}{KP}\Id_N + \frac1K\sum_{j=1}^K\frac{\Dm_j(\tau)}{1+\delta_j}\RB^{-1}.
\end{align}
Using this expression, we have
\begin{align}\nonumber
& \trace\Tm_P^{-1}\Tm_P'\\\nonumber & = -\trace\Tm_P^{-1}\Tm_P\ \frac{d}{d\, \tau}\LB\frac{L}{KP}\Id_N + \frac1K\sum_{j=1}^K\frac{\Dm_j(\tau)}{1+\delta_j}\RB\Tm_P\\\nonumber
&= -\trace\Tm_P\LB\frac1K\sum_{j=1}^K\frac{(1+\delta_j)\Dm_j'(\tau)-\delta_j'\Dm_j(\tau)}{(1+\delta_j)^2}\RB\\
&= \sum_{j=1}^K\frac{\delta_j'\delta_j - (1+\delta_j)\frac{1}{K}\trace\Dm_j'(\tau)\Tm_P}{(1+\delta_j)^2}
\end{align}
Plugging this expression into \eqref{eq:intermediate} and replacing $\delta_j$ by $\frac{1}{K}\trace\Dm_j(\tau)\Tm_P$ leads to
\begin{align}
 \overline{R}'(\tau) = \frac1N\sum_{j=1}^K \frac{\frac1K\trace\Dm'_j(\tau)\Tm_P}{1+\frac{1}{K}\trace\Dm_j(\tau)\Tm_P}.
\end{align}
In \cite[Proposition 5.3]{hachem08}, it is proved that 
\begin{align}\label{eq:bounds}
 \LB\frac{L}{KP}+\max_{i,j}\overline{v}_{ij}(\tau)\RB^{-1}\leq \LSB\Tm_P\RSB_{ii}\leq \frac{KP}{L}.
\end{align}
Since both $\overline{v}_{ij}(\tau)$ and $\overline{v}_{ij}'(\tau)$ are positive for $\tau,P>0$, it follows from \eqref{eq:bounds} that $\frac1K\trace\Dm'_j(\tau)\Tm_P>0$ and  $\frac{1}{K}\trace\Dm_j(\tau)\Tm_P>0$. This implies $\overline{R}'(\tau)>0$ which concludes the proof.

\section{Proof of Theorem~\ref{th:concavity}}\label{th:concavity_proof}
We want to show that $\overline{R}''(\tau)<0$. Under the assumption of a doubly regular variance profile matrix $\overline{\Vm}(\tau)$, the implicit matrix equation $\Tm(z)$ \eqref{eq:T} of Theorem~\ref{thm:det_equ}~(i) reduces to a scalar equation, such that $\Tm(z) = t(z)\Id_N$, where
\begin{align}
 t(z) = \frac{1}{-z+\frac{\Kc(\tau)}{1+\Kc(\tau)t(z)}}.
\end{align}
The unique solution to this equation (such that $t(z)\in\Sc$) can be given in closed-form as
\begin{align}
 t(z) = \frac{\sqrt{1-\frac{\Kc(\tau)}{z}}-1}{2\Kc(\tau)}.
\end{align}
Let $t_P=t(-\frac{L}{KP})$. By Theorem~\ref{th:derivative}, the first derivative of $\overline{R}(\tau)$ can be written as
\begin{align}
\overline{R}'(\tau) &= \frac1N\sum_{j=1}^N\frac{\frac1N\trace\Dm_j'(\tau)t_P}{1+\frac1N\trace\Dm_j(\tau)t_P} =  \frac{t_P\Kc'(\tau)}{1+ t_P\Kc(\tau)}
\end{align}
where $\Kc'(\tau)=\frac{d\,}{d\, \tau}\Kc(\tau)$.
The second derivative is given as
\begin{align}
 \overline{R}''(\tau) =  \frac{t_P'\Kc'(\tau) + t_P\Kc''(\tau)[1+ t_P\Kc(\tau)]-[t_P\Kc'(\tau)]^2}{[1+ t_P\Kc(\tau)]^2}.
\end{align}
We now need to verify that the numerator of the last equation is negative. One can easily verify from \eqref{eq:vareffd} and \eqref{eq:vard} that $\Kc'(\tau)>0$ and it follows from \eqref{eq:bounds} that $t_P>0$. It remains to check that $t_P'<0$ and $\Kc''(\tau)<0$. Write therefore $t_P$ as
\begin{align}
 t_P &= \frac{\sqrt{1+\frac{KP}{L}\Kc(\tau)}-1}{2\Kc(\tau)}= \frac{KP}{2L\LB\sqrt{1+\frac{KP}{L}\Kc(\tau)}+1\RB}
\end{align}
which is a strictly decreasing function of $\tau$ since $\Kc'(\tau)>0$. Hence, we have that $t_P'<0$. In order to show that $\Kc''(\tau)<0$, define the two auxiliary functions $
 \hat{\Kc}(\tau) = \frac1N \sum_{i=1}^N\hat{v}_{ij}(\tau)$ and $\tilde{\Kc}(\tau) = \frac1N \sum_{i=1}^N\tilde{v}_{ij}(\tau)$ which are independent of the column index $j$.
It is a simple exercise to verify that $\hat{v}_{ij}(\tau)$ are positive increasing concave functions and $\tilde{v}_{ij}(\tau)$ are positive decreasing convex functions. Due to the regularity conditions of the variance profile, one can verify from \eqref{eqn:quant_noise} that the quantization noise $\sigma_i^2$ is the same for all BS-antennas, i.e., $\sigma_i=\sigma^2$. Thus,
\begin{align}\nonumber
 \Kc(\tau)& = \frac1N \sum_{i=1}^N \overline{v}_{ij}(\tau)= \frac1N \sum_{i=1}^N \frac{\hat{v}_{ij}(\tau)}{1+\sigma^2+\frac{PN}{L}\tilde{\Kc}(\tau)}\\
&= \frac{\hat{\Kc}(\tau)}{1+\sigma^2+\frac{PN}{L}\tilde{\Kc}(\tau)}
\end{align}
Since both $\hat{\Kc}(\tau)$ and $(1+\sigma^2+\frac{PN}{L}\tilde{\Kc}(\tau))^{-1}$ are positive increasing concave functions, it follows from \cite[Exercise 3.32 (b)]{boyd_cvx} that the same holds also for their product. Hence, $\Kc''(\tau)<0$ and, thus, $\overline{R}''(\tau)<0$.

\section{Proof of Theorem~\ref{th:convergence}}\label{th:convergence_proof}
 We expand the difference $R_\text{net}(\tau^*)-R_\text{net}(\overline{\tau}^*)$ as follows:
\begin{align}\nonumber
 R_\text{net}(\tau^*)-R_\text{net}(\overline{\tau}^*) =&\ \ \ \LSB R_\text{net}(\tau^*)-\overline{R}_\text{net}(\tau^*)\RSB\\\nonumber& +\LSB \overline{R}_\text{net}(\tau^*) - \overline{R}_\text{net}(\overline{\tau}^*)\RSB\\\label{eqn:differences}& +\LSB \overline{R}_\text{net}(\overline{\tau}^*) -R_\text{net}(\overline{\tau}^*)\RSB.
\end{align}
From Theorem~\ref{thm:det_equ}~(ii), we have that the first and last term of the right-hand side (RHS) of \eqref{eqn:differences} vanish asymptotically, i.e.,
\begin{align}\label{eqn:part1}
 R_\text{net}(\tau^*)-\overline{R}_\text{net}(\tau^*)&\xrightarrow[K\rightarrow \infty]{} 0\\ \label{eqn:part11}
\overline{R}_\text{net}(\overline{\tau}^*) -R_\text{net}(\overline{\tau}^*)&\xrightarrow[K\rightarrow \infty]{} 0.
\end{align}
By the definition of $\tau^*$ and $\overline{\tau}^*$, we have for the LHS of \eqref{eqn:differences} and the second term on the RHS of \eqref{eqn:differences}
\begin{align}\label{eqn:part2}
 R_\text{net}(\tau^*)-R_\text{net}(\overline{\tau}^*) \ge 0, \quad  \overline{R}_\text{net}(\tau^*) - \overline{R}_\text{net}(\overline{\tau}^*) \le 0.
\end{align}
Equations \eqref{eqn:differences}, \eqref{eqn:part1}, \eqref{eqn:part11}, and \eqref{eqn:part2} together imply that
\begin{align}\label{eqn:conv1}
 R_\text{net}(\tau^*)-R_\text{net}(\overline{\tau}^*)&\xrightarrow[K\rightarrow \infty]{} 0\\\label{eqn:conv2}
\overline{R}_\text{net}(\tau^*) - \overline{R}_\text{net}(\overline{\tau}^*)&\xrightarrow[K\rightarrow \infty]{} 0.
\end{align}
Equation~\eqref{eqn:conv1} together with Theorem~\ref{thm:det_equ}~(ii) proofs the first part of the theorem.  
Assume now that $\overline{\Vm}(\tau)$ is a doubly regular matrix. Since $\overline{R}_\text{net}(\tau)$ is by Theorem~\ref{th:concavity} a strictly concave function which takes its unique maximum at point $\overline{\tau}^*$, \eqref{eqn:conv2} implies that $\tau^*-\overline{\tau}^*\to 0$ as $K\to\infty$.

\bibliographystyle{IEEEtran}
\bibliography{IEEEabrv,bibliography}

\begin{IEEEbiography}[{\includegraphics[width=1in,height=1.25in,clip,keepaspectratio]{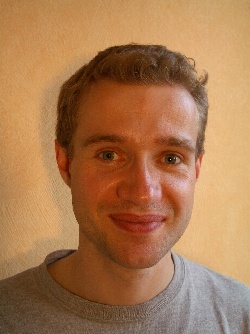}}]{Jakob Hoydis}(S'08) received the diploma degree (Dipl.-Ing.) in electrical engineering and information technology from RWTH Aachen University, Germany, in 2008. From May 2008 to April 2009, he was a research assistant at the Institute for Networked Systems, RWTH Aachen University. Since May 2009, he is working toward his Ph.D. degree in the area of cooperative communications and network MIMO at the Department of Telecommunications, Sup\'{e}lec, Gif-sur-Yvette, France.
\end{IEEEbiography}

\begin{IEEEbiography}[{\includegraphics[width=1in,height=1.25in,clip,keepaspectratio]{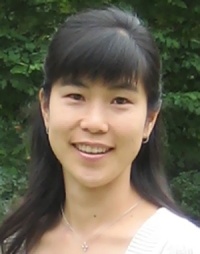}}]{Mari Kobayashi}(M'06) received the B.E. degree in electrical engineering from Keio University, Yokohama, Japan, in 1999, a M.S. degree in mobile radio, and Ph.D. degree from Ecole Nationale Sup\'{e}rieure des T\'{e}l\'{e}communications, Paris, France, in 2000 and 2005, respectively. From November 2005 to March 2007, she was a postdoc researcher at Centre Tecnol\`{o}gic de Telecomunicacions de Catalunya, Barcelona, Spain. Since May 2007, she has been an assistant professor at Sup\'{e}lec, Gif-sur-Yvette, France. Her current research interests include MIMO communication systems and multiuser communication theory.
\end{IEEEbiography}

\begin{IEEEbiography}[{\includegraphics[width=1in,height=1.25in,clip,keepaspectratio]{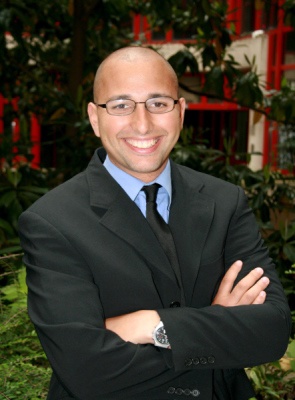}}]{M\'{e}rouane Debbah}(SM'08) received the M.Sc. and
Ph.D. degrees from the Ecole Normale Sup\'{e}rieure de Cachan, France, in 1999 and 2002, respectively. From 1999 to 2002, he worked for Motorola Labs on Wireless Local Area Networks and prospective fourth-generation systems (OFDM and MC-CDMA). From 2002 until 2003, he was appointed Senior Researcher at the Vienna Research Center for Telecommunications, Austria, working on MIMO wireless channel modeling issues. From 2003 until 2007, he was an Assistant Professor with the Mobile Communications Department of the Institute EURECOM, France. He is currently a Professor at Sup\'{e}lec, Gif-sur-Yvette, France, where he is the holder of the Alcatel-Lucent Chair on flexible radio. His research interests are in information theory, signal processing, and wireless communications.
\end{IEEEbiography}
\end{document}